\documentclass[useAMS,usenatbib]{mn2e}
\usepackage{graphics,graphicx}
\usepackage{aas_macros}
\usepackage{amssymb}
\usepackage{latexsym}
\usepackage{color}
\usepackage{epstopdf}
\usepackage{marvosym}

\begin{document}

\title[State-to-state chemistry and excitation of
  CH$^+$]{State-to-state chemistry and rotational excitation of CH$^+$
  in photon-dominated regions} \author[A.  Faure et
  al.]{A. Faure$^1$\thanks{E-mail:
    alexandre.faure@univ-grenoble-alpes.fr}, P. Halvick$^2$,
  T. Stoecklin$^2$, P. Honvault$^3$, M.~D. Ep\'ee
  Ep\'ee$^{4}$,\newauthor J.~Zs. Mezei$^{5,6,7,8}$,
  O. Motapon$^{4,9}$, I.~F. Schneider$^{5,7}$, J. Tennyson$^{10}$,
  O. Roncero$^{11}$, \newauthor N. Bulut$^{12}$,
  A. Zanchet$^{11}$\\ $^1$ Univ. Grenoble Alpes, CNRS, IPAG, F-38000
  Grenoble, France\\ $^2$ Univ. Bordeaux, CNRS, ISM, F-33400 Talence,
  France \\ $^3$ Univ. Bourgogne Franche-Comt\'e, Laboratoire ICB,
  F-21078 Dijon, France \\ $^4$ UFD Math\'ematiques, Informatique
  Appliqu\'ee et Physique Fondamentale, University of Douala,
  P. O. Box 24157, Douala, Cameroon\\ $^5$ Univ. Normandie, CNRS,
  LOMC, F-76058 Le Havre, France \\ $^6$ LSPM, Univ. Paris 13, 99
  avenue Jean-Baptiste Cl\'ement, F-93430 Villetaneuse, France\\ $^7$
  Univ. Paris-Sud, CNRS, Laboratoire Aim\'e Cotton, F-91405 Orsay,
  France\\ $^{8}$ Institute of Nuclear Research of the Hungarian
  Academy of Sciences, P.O.  Box 51, Debrecen H-4001, Hungary\\ $^{9}$
  University of Maroua, Faculty of Science, P. O. Box 814 Maroua,
  Cameroon\\ $^{10}$ Department of Physics and Astronomy, University
  College, London, Gower St., London WC1E 6BT, UK \\ $^{11}$ Instituto
  de F\'isica Fundamental, CSIC, C/ Serrano, 123, E-28006 Madrid,
  Spain\\ $^{12}$ Firat University, Department of Physics, 23169
  Elazi\~{g}, Turkey }

\date{Accepted ? Received ?}

\pagerange{\pageref{firstpage}--\pageref{lastpage}} \pubyear{2016}

\maketitle

\label{firstpage}

\begin{abstract}

We present a detailed theoretical study of the rotational excitation
of CH$^+$ due to reactive and nonreactive collisions involving
C$^+(^2P)$, H$_2$, CH$^+$, H and free electrons. Specifically, the
formation of CH$^+$ proceeds through the reaction between C$^+(^2P)$
and H$_2(\nu_{\rm H_2}=1, 2)$, while the collisional (de)excitation
and destruction of CH$^+$ is due to collisions with hydrogen atoms and
free electrons. State-to-state and initial-state-specific rate
coefficients are computed in the kinetic temperature range 10-3000~K
for the inelastic, exchange, abstraction and dissociative
recombination processes using accurate potential energy surfaces and
the best scattering methods. Good agreement, within a factor of 2, is
found between the experimental and theoretical thermal rate
coefficients, except for the reaction of CH$^+$ with H atoms at
kinetic temperatures below 50~K. The full set of collisional and
chemical data are then implemented in a radiative transfer model. Our
Non-LTE calculations confirm that the formation pumping due to
vibrationally excited H$_2$ has a substantial effect on the excitation
of CH$^+$ in photon-dominated regions. In addition, we are able to
reproduce, within error bars, the far-infrared observations of CH$^+$
toward the Orion Bar and the planetary nebula NGC~7027. Our results
further suggest that the population of $\nu_{\rm H_2}=2$ might be
significant in the photon-dominated region of NGC~7027.

\end{abstract}


\section{Introduction}

The methylidyne ion CH$^+$ was the first molecular ion to be
identified from its optical absorption spectra in the diffuse
interstellar medium (ISM) \citep{douglas41}. Since then, CH$^+$
absorption has been observed toward many background stars,
demonstrating the ubiquity of this simple carbon hydride in the
diffuse ISM. The mechanism by which it forms has remained however
elusive. Theoretical models have been indeed unable to reproduce the
large observed abundance of CH$^+$ in diffuse clouds,
[CH$^+$]/[H]$\sim 8\times 10^{-9}$ \citep[see e.g.][and references
  therein]{godard13}.

It was for a long time assumed that CH$^+$ is primarily formed by the
reaction
\begin{equation}
\rm{C^+ + H_2 \rightarrow CH^+ + H.}
\end{equation}
This reactions is barrierless but endothermic by 0.398~eV (4620~K)
\citep{hierl97}, which is much higher than the kinetic temperatures in
ordinary diffuse clouds ($T_k< 100$~K). As a result, non-equilibrium
chemistry is necessary to explain the large column densities ($\geq
10^{13}$~cm$^{-2}$) of CH$^+$ observed in the diffuse ISM. The invoked
gas heating mechanisms proposed to accelerate reaction~(1) include
C-type and magnetohydrodynamic shocks, Alfv\`en waves, turbulent
mixing and turbulent dissipation. A recent discussion of these
different scenarios in the framework of UV-dominated and
turbulence-dominated chemistries can be found in \cite{godard14}.

The vast majority of the observed CH$^+$ absorptions arise from the
ground rotational level $j=0$. The first excited level $j=1$ lies
40.1~K above the ground state and it is not sufficiently populated at
the typical density of diffuse clouds where CH$^+$ abounds ($n_{\rm
  H}\lesssim 10^2$cm$^{-3}$). The critical density of the
$j=1\rightarrow 0$ transition is indeed large, $n_{cr}\sim 4\times
10^6$cm$^{-3}$. Rotationally excited levels of CH$^+$ can be
significantly populated in the diffuse ISM only in the presence of
warm dust emission \citep{oka13}.

Emission lines from CH$^+$ are less commonly observed but they have
been detected in the visible toward the red rectangle
\citep[e.g.][]{bakker97} and in the far-infrared toward several
sources: the planetary nebula NGC~7027 \citep{cernicharo97}, the Orion
bar \citep{nagy13,parikka16}, the Orion BN/KL complex
\citep{morris16}, the massive star-forming region DR21
\citep{falgarone10}, the infrared galaxy Markarian~231
\citep{vanderwerf10} and the disk around the Herbig Be star HD~100546
\citep{thi11}. The far-infrared lines arise from pure rotational
transitions and they were detected thanks to the {\it ISO} and {\it
  Herschel} space observatories. The largest number of rotational
lines were identified in NGC~7027 and in the Orion Bar, which are two
prototypical photon-dominated regions (PDRs). In these two sources,
the rotational series $j\rightarrow j-1$ was observed from
$j=1\rightarrow 0$ at 835.137~GHz up to $j=6\rightarrow 5$ at
4976.201~GHz. In such regions the dense gas ($n_{\rm H}>
10^4$cm$^{-3}$) is illuminated by a strong far-ultraviolet (FUV)
radiation field and a reservoir of H$_2$ ro-vibrationally excited by
FUV fluorescence can provide an alternative route to overcome the
endothermicity of reaction~(1) \citep{sternberg95,agundez10}. This was
evidenced recently in the Orion Bar where a good correlation between
CH$^+$ and vibrationally excited H$_2$ was observed by
\cite{parikka16}.

Indeed, it has been shown both experimentally and theoretically that
the internal (rotational or vibrational) excitation of H$_2$ can help
to reduce or even offset the endothermicity of reaction~(1). The
rotational and vibrational energies were found to be as effective as
the translational energy in promoting the reaction
\citep{gerlich87,hierl97,zanchet13,herraez14}. As a result, the
experimental rate coefficient derived by \cite{hierl97} for the
reaction of C$^+$ with H$_2(\nu_{\rm H_2}=1)$ between 800 and 1300~K
is in the range $1-2\times 10^{-9}$~cm$^3$s$^{-1}$, in good agreement
with the Langevin limit ($1.6\times 10^{-9}$~cm$^3$s$^{-1}$). This
value represents an enhancement of about 3 orders of magnitude with
respect to the rate coefficient of ground-state H$_2(\nu_{\rm
  H_2}=0)$. This finding was confirmed theoretically by two recent
independent studies \citep{zanchet13,herraez14}. These studies were
also able to derive state-to-state rate coefficients and, in
particular, \cite{zanchet13} have provided the first rotationally
resolved rate coefficients.

State-resolved rates for the production and loss of CH$^+$ are crucial
because the rotational excitation of CH$^+$ is governed by the
competition between the radiative processes and the $j$-dependent
formation, destruction and collisional excitation processes. Indeed,
inelastic collisions with the dominant species, i.e. hydrogen atoms,
free electrons and hydrogen molecules, are not faster than the
reactive processes. Thus, in contrast to non-reactive molecules such
as CO, inelastic collisions can never fully equilibrate the rotational
populations of CH$^+$. The CH$^+$ emission spectrum is therefore
expected to retain some memory of the $j$-dependent formation process,
as discussed in \cite{black98}. As a result, when solving the coupled
equations of statistical equilibrium and radiative transfer, it is
necessary to include state-resolved formation and destruction rates in
addition to the usual radiative and inelastic rates. Strong deviations
of the level populations from local thermodynamic equilibrium (LTE)
are expected in these conditions.

Non-LTE calculations including formation and destruction rates were
recently performed to model the CH$^+$ emissions observed toward
NGC~7027, the Orion bar and Orion BN/KL
\citep{godard13,nagy13,zanchet13,morris16}. The most complete model is
that of \cite{godard13} which includes the radiative pumping of CH$^+$
vibrational and electronic states by infrared, optical and ultraviolet
photons. In all these studies, inelastic data were taken from the
state-to-state calculations of \cite{hammami09} and \cite{turpin10} on
CH$^+$-He and those of \cite{lim99} on CH$^+$-electron, and they were
complemented by extrapolations. For the destruction by reactions with
H, electrons and H$_2$, generic rates independent of $j$ were assumed,
except for CH$^+$ + H in \cite{godard13} where the
initial-state-specific rate coefficients extracted by \cite{plasil11}
for $j=0, 1, 2$ at $T_k\leq 60$~K were used. For the formation by the
reaction of C$^+$ with H$_2$, the theoretical data of \cite{zanchet13}
were employed in the most recent studies \citep{zanchet13,morris16}
while in the previous models of \cite{godard13} and \cite{nagy13} the
formation rates were expressed as a Boltzmann distribution at an
effective formation temperature. The major result of these studies is
that in warm and dense PDR conditions, the ``formation'' or
``chemical'' pumping via the reaction C$^+$+H$_2(\nu_{\rm H_2}$=1) is
the dominant source of the rotational excitation of CH$^+$($j>1$). The
use of initial-state-specific formation rates (instead of a Boltzmann
distribution) was also found to have a substantial impact on the
distribution of the highest CH$^+$ levels ($j\geq 4$)
\citep{zanchet13}. Incidentally, the density in the PDR can be one to
two orders of magnitude below the values inferred from traditional
excitation models \citep{godard13}. All the models published so far
were however hampered by the lack of initial-state-specific rate
coefficients for the destruction of CH$^+$ at $T_k>60$~K. In addition,
inelastic data for CH$^+$ + H and CH$^+$ + H$_2$ were simply scaled
from those of CH$^+$ + He, which is questionable since collisions with
H and H$_2$ are reactive.

Here we provide the first comprehensive set of theoretical
state-to-state rate coefficients for the inelastic and reactive
collisions of CH$^+$ with hydrogen atoms and free electrons in the
temperature range 10-3000~K. This data is computed from
state-of-the-art theoretical methods using the most accurate
interaction potentials. Theoretical approaches include the
time-independent quantum mechanical (TIQM) method, quasi-classical
trajectory (QCT) calculations, the $R$-matrix adiabatic nuclei
approach and the multichannel quantum defect theory (MQDT). In
addition, new quantum time-dependent (wave-packet) calculations are
performed for the formation reaction~(1) to extend the data of
\cite{zanchet13}. In the next section, the inelastic and reactive rate
coefficients are described with a brief description of the different
calculations. In Section~3, these rate coefficients are employed to
model the CH$^+$ emission spectrum for typical PDR physical
conditions. Our non-LTE model is applied, in particular, to reproduce
the CH$^+$ line fluxes observed toward the Orion Bar and the planetary
nebula NGC~7027. General conclusions are drawn in Section~4.

\section{Inelastic and reactive rate coefficients}

\subsection{Formation via C$^+$ + H$_2$}

The calculations presented in this section are an extension of the
work of \cite{zanchet13}. In those calculations, the time-dependent
wave-packet (TDWP) method was combined with the potential energy
surface (PES) developed by \cite{stoecklin05} for the electronic
ground state of CH$_2^+$. Cross sections and rate coefficients were
determined for the reaction of C$^+(^2P)$ with H$_2$ in the
ground-state $(\nu_{\rm H_2}=0, j_{\rm H_2}=0)$ and in the first
vibrationally excited state $(\nu_{\rm H_2}=1, j_{\rm H_2}=0, 1)$ for
temperatures in the range 10-5000~K. In the dynamical calculations,
only one adiabatic electronic state was thus considered, neglecting
spin-orbit couplings. In order to take them into account, it was
assumed that the cross sections for the two C$^+(^2P_{1/2})$ states
correspond to those calculated in the adiabatic approximation, while
the four C$^+(^2P_{3/2})$ states do not contribute to the
reaction. The final reaction cross sections was obtained by averaging
over all spin-orbit states, considering the proper electronic
partition function using the experimental spin-orbit splittings. The
reaction cross sections for $\nu_{\rm H_2}=0$ were found in good
agreement with the experimental results of \cite{gerlich87} while the
rate coefficients for $\nu_{\rm H_2}=1$ were shown to be a factor of
$\sim 3$ smaller than those derived by \cite{hierl97}. A very similar
result was obtained by \cite{herraez14} using a different PES and a
different methodology (QCT calculations). Part of the disagreement
between theory and experiment was attributed to the contribution of
$\nu_{\rm H_2}>1$ \citep{zanchet13} and to the effect of rotational
excitation \citep{herraez14} neglected in the experimental
derivation. Full details on the TDWP calculations can be found in
\cite{zanchet13}.

In the present work, the calculations of \cite{zanchet13} were
extended to the second vibrationally excited state of H$_2(\nu_{\rm
  H_2}=2, j_{\rm H_2}=0)$. As we will see below, the population of
$\nu_{\rm H_2}=2$ can be significant in highly FUV illuminated
regions. It should be noted that while the rotational excitation of
H$_2$ is crucial in $\nu_{\rm H_2}=0$ (because reaction~(1) is
endothermic for $j_{\rm H_2}\leq 7$), it has only a modest impact in
vibrationally excited levels $\nu_{\rm H_2}\geq 1$ for which
collisions leading to CH$^+(v=0$) display no threshold
\citep{zanchet13,herraez14}. This is illustrated in Fig.~\ref{occu}
below, where we have plotted the rotational distribution of the
nascent CH$^+$($\nu=0, j$) product from the three reactions
C$^+$+para-H$_2(\nu_{\rm H_2}=1,j_{\rm H_2}=0)$,
C$^+$+ortho-H$_2(\nu_{\rm H_2}=1,j_{\rm H_2}=1)$ and
C$^+$+para-H$_2(\nu_{\rm H_2}=2,j_{\rm H_2}=0)$ at a kinetic
temperature $T_k=500$~K. We can first observe that the rotational
state of H$_2(\nu_{\rm H_2}=1)$ (i.e. para or ortho ground-state) has
a negligible influence on the CH$^+$ distribution, as expected. We
also observe that the probability distributions for $\nu_{\rm H_2}=1$
and $\nu_{\rm H_2}=2$ peak at $j\sim6$ and $j\sim 10$,
respectively. These distributions do not follow Maxwell-Boltzmann
functions but they have, however, a resemblance to Maxwellian
distributions at $T_f=1000$~K and $T_f=5000$~K, respectively. As the
reaction enthalpy is $\sim -1500$~K for H$_2(\nu_{\rm H_2}=1,j_{\rm
  H_2}=1)$ and $\sim -7000$~K for H$_2(\nu_{\rm H_2}=2,j_{\rm
  H_2}=0)$, this suggests that about $2/3$ of the exothermicity of the
reaction is transferred to CH$^+$ rotation. Thus, although approximate
and confusing, the concept of ``formation temperature'' is not
irrelevant. We note also that the effect of kinetic temperature is
moderate in the range $T_k=10-5000$~K: the distribution broadens with
increasing temperature but the distribution peak is not shifted.

\begin{figure}
\includegraphics*[width=7.5cm,angle=-90.]{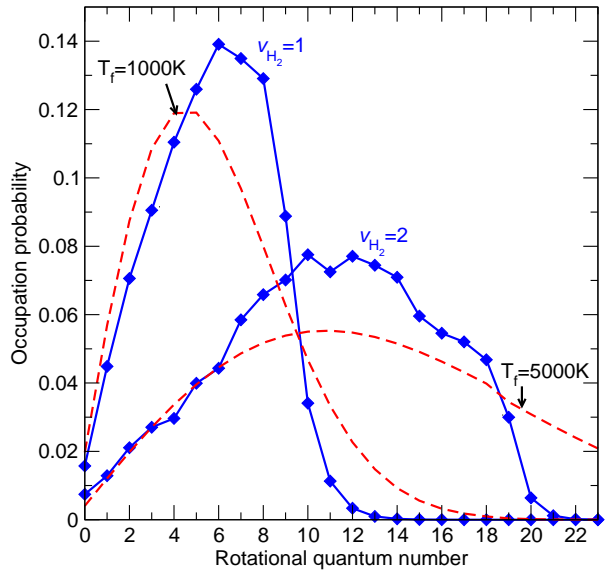}
\caption{Occupation probability of the nascent CH$^+$($v=0, j$)
  product from the reaction of C$^+$ with H$_2(\nu_{\rm H_2}=1,j_{\rm
    H_2}=1)$, H$_2(\nu_{\rm H_2}=1,j_{\rm H_2}=0)$ and H$_2(\nu_{\rm
    H_2}=2,j_{\rm H_2}=0)$ at kinetic temperature $T_k$=500~K.  The
  Boltzmann distributions at formation temperatures $T_f=1000$ and
  5000~K are plotted as dashed lines.}
\label{occu}
\end{figure}

\subsection{Excitation and destruction by hydrogen atoms}

The calculations presented in this section are an extension of the
work of \cite{werfelli15}. In those calculations, the TIQM code
\texttt{ABC} \citep{skouteris00} was combined with a new
full-dimensional PES for the ground electronic state of CH$_2^+$. The
\texttt{ABC} code was also checked for the first time against an
accurate quantum method in the case of a complex forming reaction with
a deep potential well. Cross sections and rate coefficients were
determined for CH$^+(j=0-7$) and kinetic temperatures in the range
5-800~K. The thermal rate coefficients (i.e. averaged over the
Boltzmann rotational distributions) were found in good agreement with
the experimental data in the range 50-800~K. At lower temperatures,
however, the steep fall-off observed experimentally by \cite{plasil11}
was not reproduced by the calculations of \cite{werfelli15}, in
contrast to other recent theoretical works based on different PESs
\citep{warmbier11,grozdanov13,bovino15,li15}. However, the analysis by
\cite{werfelli15} has shown that those PESs have incorrect long-range
behaviour and that the seemingly good agreement with the
low-temperature data is fortuitous. Full details on the PES and the
TIQM calculations can be found in \cite{werfelli15}.

In the present work, the calculations of \cite{werfelli15} were
extended to CH$^+(j=8-13$) and kinetic temperatures up to
$T_k=3000$~K. Levels higher than $j=13$ are not considered because
they lie above the first vibrational threshold of CH$^+$ (which opens
at $\sim$3700~K). Since TIQM calculations are computationally highly
expensive for CH$^+$ + H, we resorted here to QCT calculations using
the same PES as \cite{werfelli15}. The QCT calculations were performed
for 29 values of the collision energy $E_c$ distributed in the range
1-1600 meV, thus allowing us to calculate the rate coefficients in the
range 10-3000~K. Batches of 50,000 trajectories were computed for each
value of $E_c$ and for each initial rovibrational state ($\nu$,$j$) of
CH$^+$, with $\nu$=0 and 0 $\leq j \leq $ 13. For each batch, the
maximum impact parameter was adapted to the inelastic collisions. The
zero point energy (ZPE) leakage is a well known shortcoming of QCT
calculations, and it is even more pronounced as the number of open
rovibrational levels decreases. This failure can be corrected with the
help of the Gaussian binning (GB) procedure
\citep{bonnet2013}. However, in the case of inelastic processes, the
vibrationally adiabatic trajectories give large contributions which
overly dominate the GB probabilities \citep{bonnet2013}. Therefore, we
treated the vibrationally adiabatic trajectories with the standard
binning procedure, and use the GB procedure for the remaining
trajectories. Finally, the QCT rate coefficients were scaled in order
to provide an extension up to 3000~K of the quantum rate coefficients
previously calculated below 800~K and for 0 $\leq j \leq $ 7. For 8
$\leq j \leq $ 13, an average scaling factor was applied.

\begin{figure}
\includegraphics*[width=7.5cm,angle=-90.]{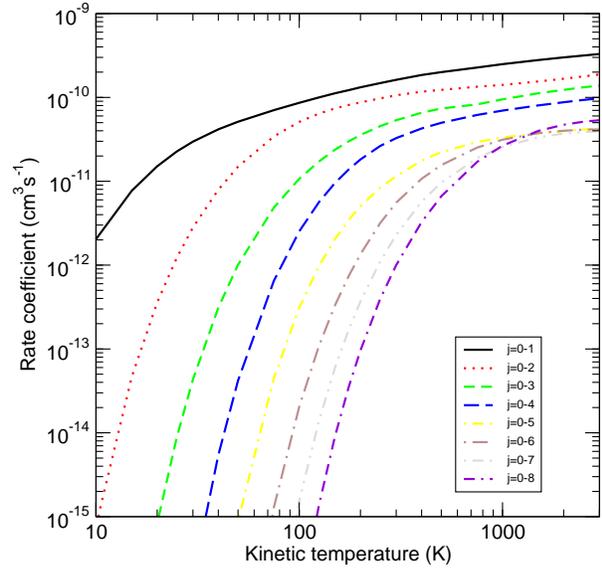}
\caption{Rate coefficients for rotational excitation of CH$^+$ out of
  the ground state $j=0$ by H as functions of the kinetic
  temperature. These data include both inelastic and exchange channels
  (see text).}
\label{ine}
\end{figure}

In Fig.~\ref{ine}, the rate coefficients for the sum of the inelastic
(nonreactive) and exchange processes CH$^+(j=0)$ + H $\rightarrow$
CH$^+(j')$+H are plotted as functions of the kinetic temperature for
$j'=1-8$. We observe a strong increase of the rate coefficients with
increasing temperature which mainly reflects the excitation thresholds
of the different transitions (e.g. $0\rightarrow 1$ opens at 40~K). As
a result, the transition with $\Delta j=1$ is largely favoured at low
temperature but this propensity holds also at high temperature,
indicating that the cross sections follow the energy gap law (at least
for $\Delta j\leq 4$). 

\begin{figure}
\includegraphics*[width=7.5cm,angle=-90.]{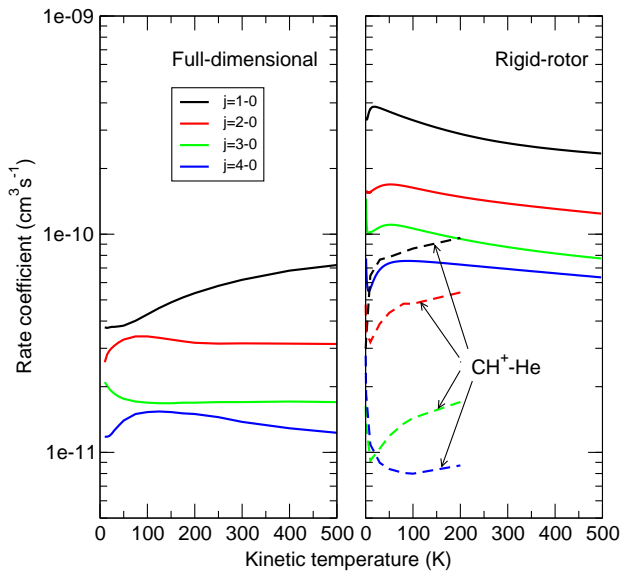}
\caption{Rate coefficients for rotational de-excitation of CH$^+$
  ($j\rightarrow 0)$ by H as functions of the kinetic
  temperature. Full-dimensional and rigid-rotor calculations are
  presented in the left and right panels, respectively. The dashed
  lines in the right panel give the results of Turpin et al. (2010)
  for CH$^+$+He.}
\label{rigid}
\end{figure}

We have also checked the reliability of the rigid-rotor approximation
for the CH$^+$ excitation by H. It was shown recently that this
approximation leads to good results in the case of the endothermic
OH$^+$+H reaction on the quartet PES whose minimum is similar to a van
der Waals well \citep{bulut15,stoecklin15}. Its accuracy was found,
however, to degrade on the doublet PES where there is a much deeper
potential well, like in CH$^+$ + H \citep{bulut15}. We have performed
rigid-rotor calculations with the TIQM code \texttt{NEWMAT}
\citep{stoecklin03} combined with the CH$_2^+$ PES of
\cite{werfelli15} where the CH$^+$ bond length was frozen at its
equilibrium geometry. As shown in Fig.~\ref{rigid}, the rigid-rotor
approximation is found to be very inaccurate for CH$^+$+H. Indeed,
although full-dimensional and rigid calculations have similar
propensity rules, the cross sections differ by up to one order of
magnitude. In the full-dimensional treatment, the largest fraction of
the scattering flux is thus directed into the reactive channel, as
expected. This illustrates the importance of full-dimensionality for
this reactive system. In the right panel, we also compare the present
rigid calculations with the rigid-rotor CH$^+$--He data of
\cite{turpin10}. Perhaps surprisingly, the agreement with the
full-dimensional CH$^+$+H calculations is much better, in particular
above 100~K. This fortuitous agreement is simply due to the much less
attractive CH$^+$--He PES.

\begin{figure}
\includegraphics*[width=7.5cm,angle=-90.]{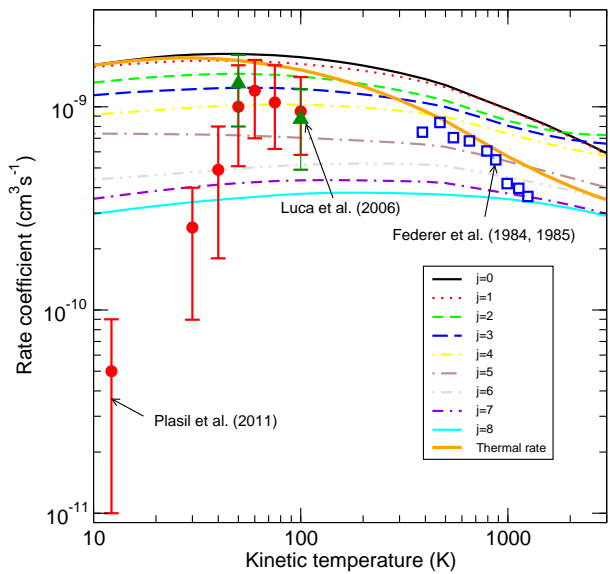}
\caption{Rate coefficients for the destruction of CH$^+$($j$) by H as
  functions of the kinetic temperature. The initial-state-specific and
  thermal rate coefficients from the present work are compared to the
  experimental data of Plasil et al. (2011), Luca et al. (2006) and
  Federer et al. (1984, 1985).}
\label{reac}
\end{figure}

In Fig.~\ref{reac}, the rate coefficients for the exothermic
destruction process CH$^+(j)$+H$\rightarrow$ C$^+$+H$_2$ are plotted
as functions of the kinetic temperature and they are compared to the
available measurements. As observed by \cite{werfelli15}, the
agreement between the calculated thermal rate coefficient and the
experimental data \citep{federer84,federer85,luca06,plasil11} is good
above $T_k$=50~K. In particular, the decrease of the rate coefficient
between 400 and 1200~K is well reproduced by the calculations. At
temperatures below 50~K, the experimental results at 12.2, 30 and
40~K) were interpreted by \cite{plasil11} as a loss of reactivity of
the lowest rotational states of CH$^+$. In contrast, we observe here a
reactivity increase with rotational cooling. This was analyzed by
\cite{werfelli15} as an increase of the probability for a nonreactive
or exchange process with increasing rotational excitation. Obviously,
further theoretical and experimental efforts are necessary to
understand this discrepancy at low temperature, as discussed in
\cite{plasil11} and \cite{werfelli15}.

\subsection{Excitation and destruction by electrons}

The calculations presented in this section are taken from
\cite{hamilton16} for the electron-impact excitation of CH$^+$ and
from Ep\'ee Ep\'ee et al. (in preparation) for the dissociative
recombination (DR) of CH$^+$ with electrons. In the calculations of
Hamilton et al. (2016), the molecular {\bf R}-matrix theory was
combined with the adiabatic nuclei rotation (ANR)
approximation. Inelastic cross sections and rate coefficients were
determined for CH$^+(j=0-11$) and kinetic temperatures in the range
1-3000~K. No experimental inelastic data is available for this system
but we note that the same methodology was applied with success to
HD$^+$ for which theoretical rate coefficients were found comparable
to within 30\% with those resulting from a fit of experimental cooling
curves \citep{shafir09}. Very good agreement with the MQDT
calculations of \cite{motapon14} for HD$^+$ and those of \cite{epee16}
for H$_2^+$ was also observed. In the calculations of Ep\'ee Ep\'ee et
al. (in preparation), the MQDT formalism was employed: based on the
diabatic potential energy curves and the interaction between the
ionization and the dissociation continua within the $^2\Pi$ symmetry
used by \cite{carata00}, we have computed the interaction, reaction
and scattering matrices, and produced the DR cross sections for the 11
lowest rotational levels of CH$^+$ in its ground electronic and
vibrational state. Rate coefficients were determined for the DR of
CH$^+(j=0-10$) at kinetic temperatures between 10 and 3000~K. The
CH$^+$ thermal DR cross section (after convolution with the
experimental resolution) was found in satisfactory agreement with the
measurements of \cite{amitay96} below 100~meV, and especially below
50~meV. Full details will be published elsewhere (Ep\'ee Ep\'ee et
al., in preparation).

The rate coefficients for the inelastic process CH$^+(j=0)$ + e$^-$
$\rightarrow$ CH$^+(j')$+ e$^-$ are discussed in Hamilton et
al. (2016) where full details can be found. In their Fig.~3, the
temperature dependences of the rate coefficients are found to be
similar to those observed here for CH$^+$ + H (Fig.~\ref{ine}) due to
the threshold effects. Dipolar transitions $\Delta j=1$ were found to
be preferred, as expected for a strongly polar target. We note that
this is in contrast with the previous {\bf R}-matrix results of Lim et
al. (1999) who found the cross sections for the $\Delta j=2$
transitions to be greater than $\Delta j=1$ transitions. This
difference was attributed by Hamilton et al. (2016) to the improved
treatment of polarization in the new calculations.

\begin{figure}
\includegraphics*[width=7.5cm,angle=-90.]{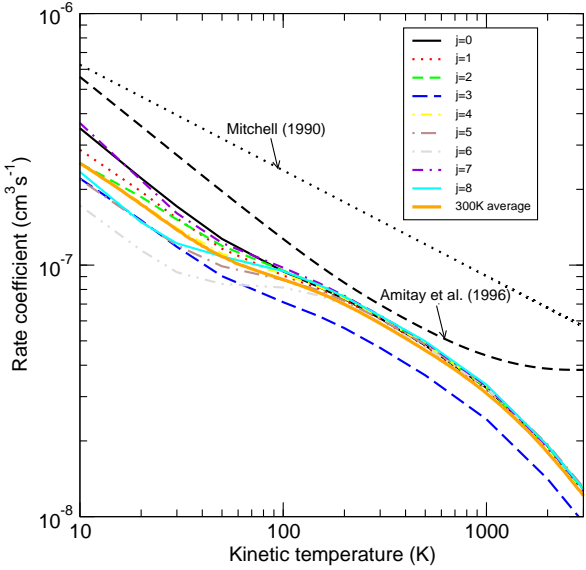}
\caption{Rate coefficients for the destruction of CH$^+$($j$) by
  electrons as functions of the kinetic temperature. The
  initial-state-specific and 300~K thermal rate coefficients from the
  present work are compared to the experimental results of Amitay et
  al. (1996) and Mitchell (1990).}
\label{e-dr}
\end{figure}

In Fig.~\ref{e-dr}, the initial-state-specific DR rate coefficients
are plotted as functions of the kinetic temperature. These rate
coefficients were obtained by averaging the cross sections over
isotropic Maxwell-Boltzmann velocity distributions. The thermal
average at a rotational temperature of 300~K is also shown and it is
compared to two sets of experimental rate coefficients. The first was
obtained from a thermal average of the DR cross sections measured by
\cite{amitay96}, where the ions were assumed to be thermalized at the
ambient temperature of the storage ring (300~K). The second is the
experimental recommendation of \cite{mitchell90}. This later is based
on the merged-beam data of \cite{mitchell78} (divided by 2 to correct
for a calibration error). The recommendation of \cite{mitchell90} is
found to exceed the results of \cite{amitay96} by about a factor of
2. The theoretical thermal average at 300~K is significantly lower but
the agreement with \cite{amitay96} is within a factor of 2 up to
$\sim$ 1000~K. It should be noted that the variation of the DR rate
coefficients is not monotonic with $j$, i.e. there is no particular
trend with increasing $j$. Such variations were already observed in
the case of HD$^+$ \citep{motapon14} and of H$_2^+$ \citep{epee16}. In
contrast to these systems, however, the rotational effects are weak in
CH$^+$ and all initial-state-specific rates agree to within a factor
of 2.

Finally, we note that all above collisional data are available upon
request to the authors.

\section{Radiative transfer model}

In order to illustrate the impact of the newly computed rate
coefficients for the CH$^+$ formation, excitation and destruction, we
now implement these data in a non-LTE radiative transfer model of
CH$^+$ excitation in PDR conditions. PDRs are predominantly neutral
regions of the interstellar medium where the gas is exposed to FUV
radiation fields ($\sim 6-13.6$~eV). PDRs typically occur at the
boundaries of planetary nebulae, on the edges of molecular clouds and
in the nuclei of starburst and active galaxies \citep{sternberg95}. A
PDR thus starts at the fully ionized ``HII'' region where only
far-ultraviolet radiation penetrates the neutral gas, i.e. stellar
photons that cannot ionize hydrogen atoms but do ionize elements with
low ionization potential ($<$ 13.6~eV ) such as carbon. The C$^+$
column density thus increases with the intensity of FUV radiation and
it decreases with gas density. Inside the PDR, the FUV photon flux is
indeed limited by H$_2$ and dust absorption. Beyond the edge of the
C$^+$ zone, carbon is rapidly incorporated into CO. PDR models
indicate that the CH$^+$ abundance peaks at visual extinctions
$A_V\sim 0.1-1$ where the carbon is fully ionized and where the amount
of vibrationally excited H$_2$ is high ($f^*=n(\nu_{\rm H_2}
>0)/n(\nu_{\rm H_2}=0)> 10^{-6}$)
\citep{agundez10,godard13,nagy13}. In this zone, hydrogen is
predominantly in atomic form, most of the electrons are provided by
the ionization of carbon atom, the kinetic temperature $T_k$ is a few
hundreds of Kelvin and the thermal pressure ($P/k_B=n_{\rm H}T_k$) is
about $10^8$~K cm$^{-3}$. Obviously, the physical conditions in PDRs
encompass a range of temperatures and densities with a rather complex
morphology, as revealed recently with ALMA \citep{goicoechea16}. In
the following, however, we will assume that CH$^+$ probes a region
with homogeneous density and temperature, corresponding to the ``hot
gas at average density'' described by \cite{nagy16} for the Orion Bar.

As explained in the Introduction, since CH$^+$ is destroyed by
hydrogen atoms and electrons on a similar time scale as it is
rotationally equilibrated (by the same colliders), the chemical
formation and destruction rates need to be included when computing the
statistical equilibrium equation \citep[e.g.][]{vandertak07}:
\begin{equation}
\frac{dn_i}{dt}=\sum_{i\ne j}^N n_jP_{ji}-n_i\sum_{i\ne
  j}^NP_{ij} + F_i-n_iD_i=0.
\label{stat}
\end{equation}
In this equation, $N$ is the number of levels considered, $n_i$ is the
level population of level $i$, $P_{ji}$ and $P_{ij}$ are the
populating and depopulating rates:
\begin{equation}
P_{ij}= \left\lbrace
\begin{array}{ccc}
  A_{ij}+B_{ij}\bar{J_{\nu}}+C_{ij}  & (i>j) \\
  B_{ij}\bar{J_{\nu}}+C_{ij}  & (i<j), 
\end{array}\right.
\label{rates}
\end{equation}
and $F_i$ and $D_i$ are the state-resolved formation and destruction
rates, respectively, of level $n_i$. In Eq.~\ref{rates},
$\bar{J_{\nu}}$ denotes the specific intensity integrated over line
profile and solid angle and averaged over all directions. The
proportionality rates $A_{ij}$ and $B_{ij}$ are the Einstein
coefficients for spontaneous and stimulated emission, respectively,
and the $C_{ij}$ are the collisional rates, i.e. the product of the
collisional rate coefficients (in cm$^3$s$^{-1}$) and the collider
density (in cm$^{-3}$), summed over all possible collision partners
(here H and e$^-$). Note that in Eq.~\ref{rates}, the notation $i>j$
means all states $i$ with an energy higher than the energy of level
$j$.

The main difficulty in solving the radiative transfer problem is the
interdependence between the level populations and the local radiation
field. Among approximate methods, the escape probability technique
\cite[e.g.][]{castor70} is widely employed. It is not adapted to model
inhomogeneous sources, for which more sophisticated treatments are
available \cite[see e.g.][and references therein]{lambert15}, but it
is very useful in describing global average properties. In the present
work, we have employed the public version of the \texttt{RADEX}
code\footnote{http://home.strw.leidenuniv.nl/$\sim$moldata/radex.html}
which uses the escape probability formulation assuming an isothermal
and homogeneous medium. In its public version, the formation and
destruction rates are assumed to be zero. We have therefore
implemented in \texttt{RADEX} the inclusion of the source ($F_i$ in
cm$^3$s$^{-1}$) and sink ($D_i$ in s$^{-1}$) terms, which requires
only minor modifications. It should be noted that only the relative
values of $F_i$ matter since the CH$^+$ column density is fixed within
\texttt{RADEX}. As a result, the column density of C$^+$ and
H$_2(\nu_{\rm H_2}>0)$ are implicit parameters in our
calculations. The adopted formation rates correspond to H$_2$ in the
inital state ($\nu_{\rm H_2}=1, j_{\rm H_2}=1$) or, alternatively,
($\nu_{\rm H_2}=2, j_{\rm H_2}=0$) (see below).

\subsection{Prototypical PDR}

The input parameters to \texttt{RADEX} are the kinetic temperature,
$T_k$, the column density of CH$^+$, $N({\rm CH^+})$, the line width,
$\Delta v$, and the density of the colliding partners, $n_{\rm H}$ and
$n({e^-})$. For a prototypical PDR, the thermal pressure is of the
order of $10^8$~K cm$^{-3}$ (see above). Here the density was set at
$n_{\rm H}=2\times 10^5$~cm$^{-3}$ and the kinetic temperature at
$T_k=500$~K. We adopted a typical electron fraction of
$x_e=n(e^-)/n_{\rm H}= 10^{-4}$, as expected if carbon is fully
ionized (for a standard ISM carbon elemental abundance). The line
width was fixed at 5~kms$^{-1}$ as observed in the Orion Bar
\citep{nagy13} and the CH$^+$ column density was set at
10$^{14}$~cm$^{-2}$, corresponding to an average CH$^+$ abondance of
$\sim 10^{-8}$. Finally, we assumed the cosmic microwave background
(CMB) as the only background radiation field with a temperature of
2.73~K. It was indeed shown by \cite{godard13} that radiative pumping
has only a minor influence on the CH$^+$ rotational distribution.

\begin{figure}
\includegraphics*[width=7.5cm,angle=-90.]{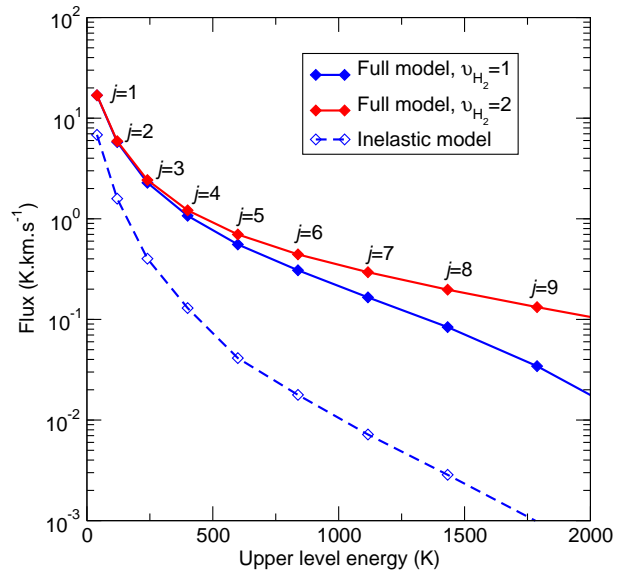}
\caption{CH$^+$ line fluxes of $j\rightarrow j-1$ rotational
  transitions as functions of the upper level energy as predicted by
  our non-LTE calculations for the following PDR conditions: a
  pressure of 10$^8$~K cm$^{-3}$ ($T_k=500$~K, $n_{\rm H}=2\times
  10^5$~cm$^{-3}$), an electron fraction $x_e$ of 10$^{-4}$ and a
  CH$^+$ column density of $10^{14}$~cm$^{-2}$. See text for details
  on the three different models.}
\label{fref}
\end{figure}

In Fig.~\ref{fref}, the line fluxes of the rotational series
$j\rightarrow j-1$ from $j=1\rightarrow 0$ to $j=9 \rightarrow 8$ are
plotted as functions of the upper level energies for three different
calculations: in the first, the public version of \texttt{RADEX} was
employed, which entirely neglects the formation and destruction rates
of CH$^+$. The rotational distribution of CH$^+$ is therefore
established only through inelastic collisions with hydrogen atoms and
free electrons. In the second calculation, the full model, which
considers the formation of CH$^+$ {\it via} C$^+$+H$_2(\nu_{\rm
  H_2}=1, j_{\rm H_2}=1)$ and its destruction {\it via} reactive
collisions with hydrogen atoms and electrons, is employed. In the
third calculation, the formation rates are those corresponding to
H$_2(\nu_{\rm H_2}=2, j_{\rm H2}=0)$. The relative populations of the
vibrationally excited states of H$_2$ are not known with precision but
the state $\nu_{\rm H_2}=2$ can reach a population of $\sim 0.3-0.5$
with respect to the $\nu_{\rm H_2}=1$ state, e.g. in the Orion bar
\citep{vanderwerf96,walmsley00}. The $\nu_{\rm H_2}=2$ state can
therefore play an important role in the chemical pumping of CH$^+$. We
can first observe in Fig.~\ref{fref} that the inclusion of formation
and destruction rates (``full model'') increases the fluxes of the
lines by at least a factor of 2 and up to two orders of
magnitude. This clearly illustrates the importance of chemical
pumping, even for the lowest $j=1\rightarrow 0$ transition. The impact
of the $\nu_{\rm H_2}=2$ state is to further increase the line fluxes,
especially for transitions above $j=4$. This is easily explained by
looking at the rotational distributions of the nascent CH$^+(\nu=0,
j)$ in Fig.~\ref{occu}: the probability for $\nu_{\rm H_2}=2$ peaks at
higher $j$ than $\nu_{\rm H_2}=1$ and this initial hotter distribution
translates into a larger final excitation. This demonstrates that the
memory of the CH$^+$ formation process is transferred to, and
preserved in the rotational distribution. In other words, the
radiative cascade subsequent to the initial formation of CH$^+$ is
dominant over the pure inelastic (excitation and de-excitation)
processes.

\subsection{The Orion Bar}

The Orion Bar PDR is the archetypal edge-on molecular cloud surface
illuminated by FUV radiation from nearby massive stars. The presence
of vibrationally excited H$_2$ in this source is supported by infrared
observations of the $\nu_{\rm H_2}=1\rightarrow 0$ and $\nu_{\rm
  H_2}=2\rightarrow 1$ transitions \citep{vanderwerf96}. CH$^+$ was
first detected in the Orion Bar by \cite{naylor10} and
\cite{habart10}, based on {\it Herschel}-SPIRE maps of the
$j=1\rightarrow 0$ transition. These studies were then extended thanks
to {\it Herschel}-HIFI and {\it Herschel}-PACS data with the detection
of the full rotational series up to $j=6\rightarrow 5$ by
\cite{nagy13}. Recently, the spatial distribution of CH$^+$ was found
to be well correlated with that of the $\nu_{\rm H_2}=1\rightarrow 0$
line of H$_2$ in this source \citep{parikka16}, as predicted by
\cite{agundez10}. Previous non-LTE models including chemical pumping
for the Orion Bar can be found in \cite{godard13}, \cite{zanchet13}
and \cite{nagy13}.

\begin{figure}
\includegraphics*[width=7.5cm,angle=-90.]{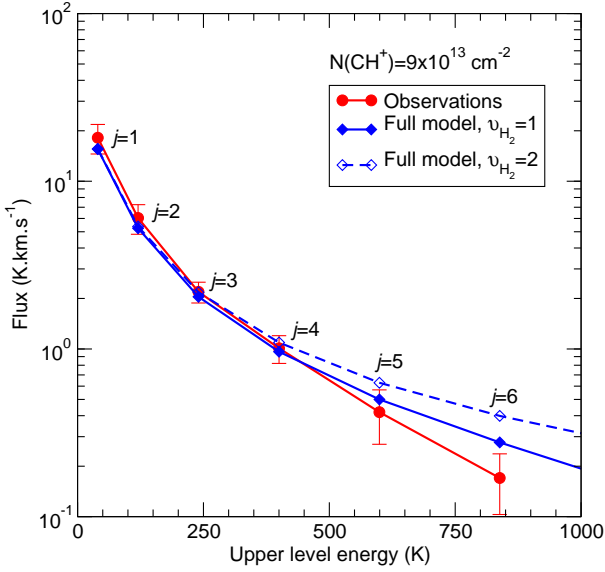}
\caption{Same as in Fig.~7 except that the CH$^+$ column density is
  adjusted to best reproduce the observations of Nagy et al. (2013)
  toward the Orion bar.}
\label{fori}
\end{figure}

In our non-LTE calculations, we adopted the same physical conditions
as for the above prototypical PDR, except that the CH$^+$ column
density is a free parameter adjusted to best reproduce the
observations of \cite{nagy13}, assuming a unit filling
factor\footnote{We note that the flux of the two {\it Herschel}-HIFI
  lines reported in Table~2 of Nagy et al. (2013) were corrected for
  main beam efficiencies (Nagy, Z. private communication).}. Very good
agreement is observed in Fig.~\ref{fori} between the model ($\nu_{\rm
  H_2}=1$) and the observations for a CH$^+$ column density of
$9\times 10^{13}$~cm$^{-2}$, corresponding to a column density per
unit line width of $1.8\times
10^{13}$~cm$^{-2}$(km~s$^{-1})^{-1}$. Indeed, the calculations agree
within error bars with {\it Herschel} data, except for the highest
transition. This column density is in excellent agreement with that
derived by \cite{morris16} for the Orion~BN/KL average. It is however
a factor of 10 lower than the value of \cite{nagy13} for the Orion
Bar. These authors have employed similar physical conditions but lower
densities and different collisional and chemical rates, which likely
explains the differences and illustrates the importance of accurate
microphysics data. In Fig.~\ref{fori}, we have also reported our
results when the formation rates are those for H$_2(\nu_{\rm H_2}=2,
j_{\rm H2}=0)$. As expected, these rates produce a hotter rotational
distribution (see also Fig.~\ref{fref}) and the flux of the lines
$j=5\rightarrow 4$ and $j=6\rightarrow 5$ is higher than that
observed, suggesting that the relative population of $\nu_{\rm H_2}=2$
is not large in the Orion Bar. We note, finally, that the contribution
of electron collisions is small at an electron fraction $x_e=10^{-4}$:
the largest effect is a 11\% increase in the intensity of the
ground-state $j=1\rightarrow 0$ transition, as observed by
\cite{nagy13}. The excitation and destruction of CH$^+$ in this source
is therefore dominated by hydrogen collisions.

\subsection{NGC~7027}

NGC~7027 is a prototypical young planetary nebula in which the
circumstellar gas is exposed to a high FUV flux emanating from the hot
central white dwarf. A large fraction of vibrationally excited H$_2$
($f^*\sim 10^{-3}$) is predicted for this source
\citep{agundez10}. The presence of excited H$_2$ was established by
      {\it ISO} observations of several $\nu_{\rm H_2}=1\rightarrow 0$
      transitions \citep{bernard01}. CH$^+$ was discovered in this
      source by \cite{cernicharo97} with the {\it ISO} detection of
      five rotational lines from $j=2\rightarrow 1$ to $j=6\rightarrow
      5$. The ground-state line $j=1\rightarrow 0$ was detected more
      recently with {\it Herschel}-SPIRE by \cite{wesson10}. These
      authors have shown that the six rotational lines cannot be
      fitted by a single excitation temperature. The first non-LTE
      model including chemical pumping for NGC~7027 was presented by
      \cite{godard13}.

\begin{figure}
\includegraphics*[width=7.5cm,angle=-90.]{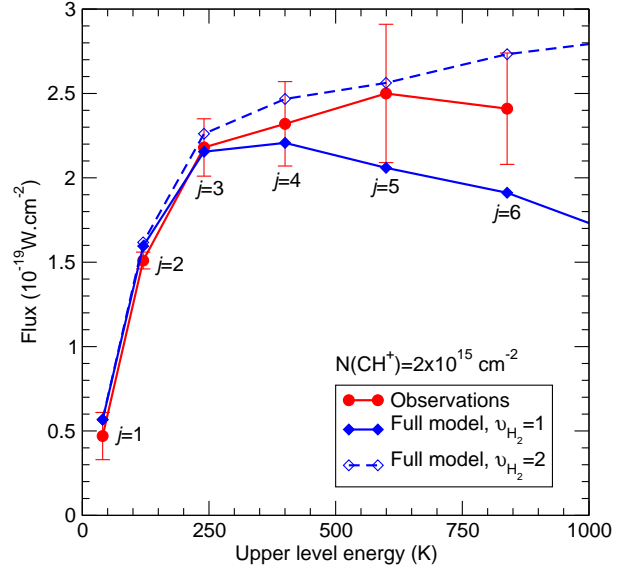}
\caption{Same as in Fig.~7 except that the electron fraction is
  $x_e=10^{-3}$ and the CH$^+$ column density is adjusted to best
  reproduce the observations of Cernicharo et al. (1997) and Wesson
  al. (2010) toward NGC~7027. Two sets of formation rates are
  employed, corresponding to either H$_2(\nu_{\rm H_2}=1)$ or
  H$_2(\nu_{\rm H_2}=2)$.}
\label{fngc}
\end{figure}

In our non-LTE calculations, we adopted the same hydrogen density and
kinetic temperature as for the above prototypical PDR but the other
parameters were modified as follows: the electron fraction was
increased to $x_e=10^{-3}$ to account for the higher elemental carbon
abundance in this carbon rich circumstellar envelope \citep{godard13},
and the line width was set at 30~km~s$^{-1}$, as in
\cite{cernicharo97}. We also assumed that the size of the CH$^+$
emission in the PDR is 10$''$ \citep{cox02}. Finally, the CH$^+$
column density is again the free parameter adjusted to best reproduce
the {\it ISO} and {\it Herschel} observations. Note that the flux
toward NGC~7027 was measured in W.cm$^{-2}$ (not in K.km.s$^{-1}$) so
that it does not decline monotically with increasing upper level
energy.  Good agreement is observed in Fig.~\ref{fngc} between the
model and the observations, although the fluxes of the highest
transitions $j=5\rightarrow 4$ and $j=6\rightarrow 5$ are
underproduced when H$_2$ is initially in $\nu_{\rm H_2}=1$. The best
model is obtained for a large CH$^+$ column density of $2\times
10^{15}$~cm$^{-2}$, corresponding to a column density per unit line
width of $6.7\times 10^{13}$~cm$^{-2}$(km s$^{-1})^{-1}$, in good
agreement with \cite{godard13}.  Fig.~\ref{fngc} also reports our
results when the formation rates are those for H$_2(\nu_{\rm H_2}=2,
j_{\rm H2}=0)$. The agreement with {\it ISO} observations is now
within error bars. This suggests that the relative population of the
state $\nu_{\rm H_2}=2$ could be large in the PDR of NGC~7027. The
observations of higher frequency transitions of CH$^+$
($j=7\rightarrow 6$, etc.) might help to confirm this result. Finally,
we note that the contribution of electron collisions is substantial
here at an electron fraction $x_e=10^{-3}$cm$^{-3}$: a 75\% increase
in the intensity of the ground-state $j=1\rightarrow 0$ transition was
observed. This effect decreases for higher transitions but electrons
still contribute above 10\% for the $j=4\rightarrow 3$ transition. As
a result, electrons compete with hydrogen atoms for the excitation and
destruction of CH$^+$ in this carbon-rich circumstellar envelope.

\section{Conclusion}

We have presented a detailed theoretical study of the rotational
excitation of CH$^+$ due to chemical pumping, excitation and
destruction. The investigated colliders were hydrogen atoms and free
electrons. State-to-state and initial-state-specific rate coefficients
were computed for the inelastic, exchange, abstraction and
dissociative recombination processes using the best available
potential energy surfaces and scattering methods. State-to-state rate
coefficients were also computed for the formation reaction,
C$^+(^2P)$+H$_2(\nu_{\rm H_2}, j_{\rm H_2})\rightarrow $CH$^+$+H by
extending the calculations of \cite{zanchet13} to the vibrationally
excited state $(\nu_{\rm H_2}=2, j_{\rm H_2}=0)$. Good agreement with
available experimental thermal rate coefficients was observed, except
for the abstraction reaction CH$^+$+H at low temperature
($<50$~K). The full set of collisional and chemical data were then
implemented in a radiative transfer model based on the escape
probability formalism. These non-LTE calculations have confirmed
previous studies that suggested that chemical pumping has a
substantial effect on the excitation of CH$^+$ in PDRs
\citep{godard13,nagy13,zanchet13}. However, in contrast to these
previous works, we have employed for the first time a comprehensive
theoretical set of fully state-to-state data. Our non-LTE model was
applied to typical PDR conditions and, in particular, to two
prototypical sources: the Orion Bar and the planetary nebula
NGC~7027. We were able to reproduce, within error bars, the {\it ISO}
and {\it Herschel} measurements by adjusting the CH$^+$ column density
as a single free parameter.

Obviously, there is no unique solution and this work can be further
improved e.g. by computing self-consistently the CH$^+$ abundance,
instead of fixing its column density. The impact of the H$_2$
rotational distribution within the different vibrational manifolds
should be also investigated. This will require to extend the
calculations on C$^+(^2P)$+H$_2(\nu_{\rm H_2}, j_{\rm H_2})$ to higher
rotational levels $j_{\rm H_2}>1$, of particular importance in the
ground vibrational state $\nu_{\rm H_2}=0$. Collisions between CH$^+$
and H$_2$ should be also considered. We have assumed in the present
work that hydrogen is in atomic form in the region of maximum CH$^+$
abundance. However, a non-negligible molecular hydrogen fraction is
necessary to form CH$^+$ from C$^+$+H$_2$. The knowledge of
state-to-state rate coefficients for the CH$^+$+H$_2$ collisions might
even provide constraints to the H$_2$ fraction. Calculations of the
PES for the electronic ground state of CH$_3^+$ are in progress in
Bordeaux (Halvick and co-workers). Finally, the generalization of this
work to the excitation of other reactive species such as OH$^+$ and
SH$^+$, for which state-resolved formation rates are becoming
available \citep{gomez14,zanchet16}, will be presented in future
works.

\section*{Acknowledgements}

This work has been supported by the Agence Nationale de la Recherche
(ANR-HYDRIDES), contract ANR-12-BS05-0011-01 And by the CNRS national
program ``Physico-Chimie du Milieu Interstellaire''. I.S acknowledges
generous financial support from R\'egion Haute-Normandie, the GRR
Electronique, Energie et Mat\'eriaux, the ``F\'ed\'eration de
Recherche Energie, Propulsion, Environnement'', the fund Bioengine,
and the LabEx EMC, via the project PicoLIBS. O.R. and A.Z. acknowledge
the financial support of Ministerio de Economia e Innovaci\'on under
grant FIS2014-52172-C2 and the European Research Council under ERC
Grant Agreement n. 610256 (NANOCOSMOS). Fran\c{c}ois Lique and Pierre
Hily-Blant are acknowledged for useful discussions.

\bibliographystyle{mn2e}

\bibliography{faure2703}

\bsp

\label{lastpage}

\end{document}